\documentclass{aa}  
\usepackage{graphicx}
\usepackage{natbib}
\usepackage{subfig}
\usepackage{epsfig} 
\usepackage{ifpdf} 
\usepackage{fancyvrb}
\usepackage{txfonts}
\usepackage{xcolor}

\begin{document} 

\title{The stellar wind velocity field of HD 77581}

\author{A. Manousakis \inst{1}\and R. Walter\inst{2,3}}
          
\institute{Centrum Astronomiczne im. M. Kopernika, Bartycka 18, PL-00716 Warszawa, Poland \\ \email{antonism@camk.edu.pl} 
\and ISDC Data Center for Astrophysics, Universit\'e de Gen\`eve, Chemin d'Ecogia 16, CH-1290 Versoix, Switzerland  
\and Observatoire de Gen\`eve, Universit\'e de Gen\`eve,  Chemin des Maillettes 51, CH-1290 Versoix, Switzerland        
}

 
  \abstract
   {}
   {The early acceleration of stellar winds in massive stars is poorly constrained. The scattering of hard X-ray photons emitted by the pulsar in the high-mass X-ray binary Vela X-1 can be used to probe the stellar wind velocity and density profile close to the surface of its supergiant companion HD 77581.
   }
   {We built a high signal-to-noise and high resolution hard X-ray lightcurve of Vela X-1 measured by Swift/BAT over 300 orbital periods of the system and compared it with the predictions of a grid of hydrodynamic simulations.}
   {We obtain a very good agreement between observations and simulations for a narrow set of parameters, implying that the wind velocity close to the stellar surface is twice larger than usually assumed with the standard beta law. Locally a velocity gradient of $\beta\sim0.5$ is favoured. Even if still incomplete, hydrodynamic simulations are successfully reproducing several observational properties of Vela X-1.}
   {}

\maketitle

\section{Introduction}

The density and velocity of stellar winds in massive stars are poorly constrained close to the stellar surface. Eclipsing high-mass X-ray binaries with short orbital orbits allow to probe the conditions of the inner stellar wind in situ. Thanks to its high X-ray flux, Vela X-1 is a perfect laboratory, with a pulsar accreting and photo-ionizing the wind of its massive (B 0.5 Ib) companion. The X-ray emission produced close to the neutron star traces instabilities of the accretion flow on large scales, which can be studied thanks to the variability of the X-ray flux and absorbing column density \citep{Walter07Winds}. 

The neutron star of Vela X-1 weights M$_{NS}=1.86$ M$_{\odot}$ \citep{Quaintrell_et_al03} and orbits its massive companion with a period of 8.96 days in an almost circular orbit \citep[$\alpha=$1.76 $R_{*}$, $e\approx 0.09$;][]{1997ApJS..113..367B}. The wind from its massive stellar companion  is characterized by a mass-loss rate of $\sim 4\times10^{-6}$ M$_{\odot}$ yr$^{-1}$ \citep{1986PASJ...38..547N} and a wind terminal velocity of $\upsilon_{\infty}\approx 1700$ km s$^{-1}$ \citep{1980ApJ...238..969D}, translating to a local wind velocity of 400 km/s at 1.2 stellar radius, assuming the commonly used $\beta=0.5$ velocity gradient.
The typical X-ray luminosity is of the order of $\sim 4 \times 10^{36}$ erg s$^{-1}$, although high variability can be observed \citep{1999A&A...341..141K}.

In this paper we present the hard X-ray lightcurve of Vela X-1, folded along the orbit, measured with an unprecedented signal to noise thanks to almost eight years of {\it Swift/BAT} observations, and compare it with the predictions of 2-D hydrodynamic simulations \citep{Manousakis_offstates}. This provides some new insight on the inner stellar wind of HD 77581 and in particular on the $\beta$ velocity law. The data and the simulations are described in  Sect. \ref{sec:DATAanalysis} and \ref{sec:hydro}, respectively. The comparison between data and simulation is presented in Sect. \ref{sec:hydro} and discussed in  Sect. \ref{sec:disc}. Section \ref{sec:conclude} summarizes our results.

\section{Swift/BAT Data and Analysis} \label{sec:DATAanalysis}

The wide field of view of the Burst Alert Telescope \citep[BAT,][]{0067-0049-209-1-14} on board of the Swift satellite allows to monitor the complete sky at hard X-rays every few hours. For sources as bright as Vela X-1, lightcurves can be obtained in different energy bands with a resolution of 1000 sec for almost a complete decade.

The Swift/BAT reduction pipeline is described in \cite{2010ApJS..186..378T} and \cite{2013ApJS..207...19B}. Our pipeline is based on the BAT analysis software HEASOFT v 6.13. A first analysis was performed with the task \texttt{batsurvey} to create sky images in the 8 standard energy bands using an input catalogue of 86 bright sources (that have the potential to be detected in single pointings) for image cleaning. Background images were then derived removing all detected excesses with the task \texttt{batclean} and weighted averaged on a daily basis. The variability of the background was then smoothed pixel-by-pixel using a polynomial model with an order equal to the number of months in the data set. The BAT image analysis was then run again using these averaged background maps. The image data were stored in a database organised by sky pixel (using a fixed sky pixel grid) by properly projecting the images on the sky pixels, preserving fluxes. This database can then be used to build images, spectra or lightcurves for any sky position.

The result of our processing was compared to the standard results presented by the Swift team (lightcurves and spectra of bright sources from the Swift/BAT 70-months survey catalogue\footnote{\texttt{http://swift.gsfc.nasa.gov/results/bs70mon/}}) and a very good agreement was found.

The Swift/BAT lightcurves of Vela X-1 were built in several energy bands. For each time bin and energy band a weighted mosaic of the selected data is first produced and the source flux is extracted assuming fixed source position and shape of the point spread function. The source signal to noise varies regularly because of intrinsic variability, its position in the BAT field of view and distance to the Sun. The lightcurves span from $\sim$ 53370 to $\sim$ 56290 MJD, i.e. over 300 orbital periods or almost 9 years.
 
\begin{figure}
\centering 
\includegraphics[width=0.5\textwidth]{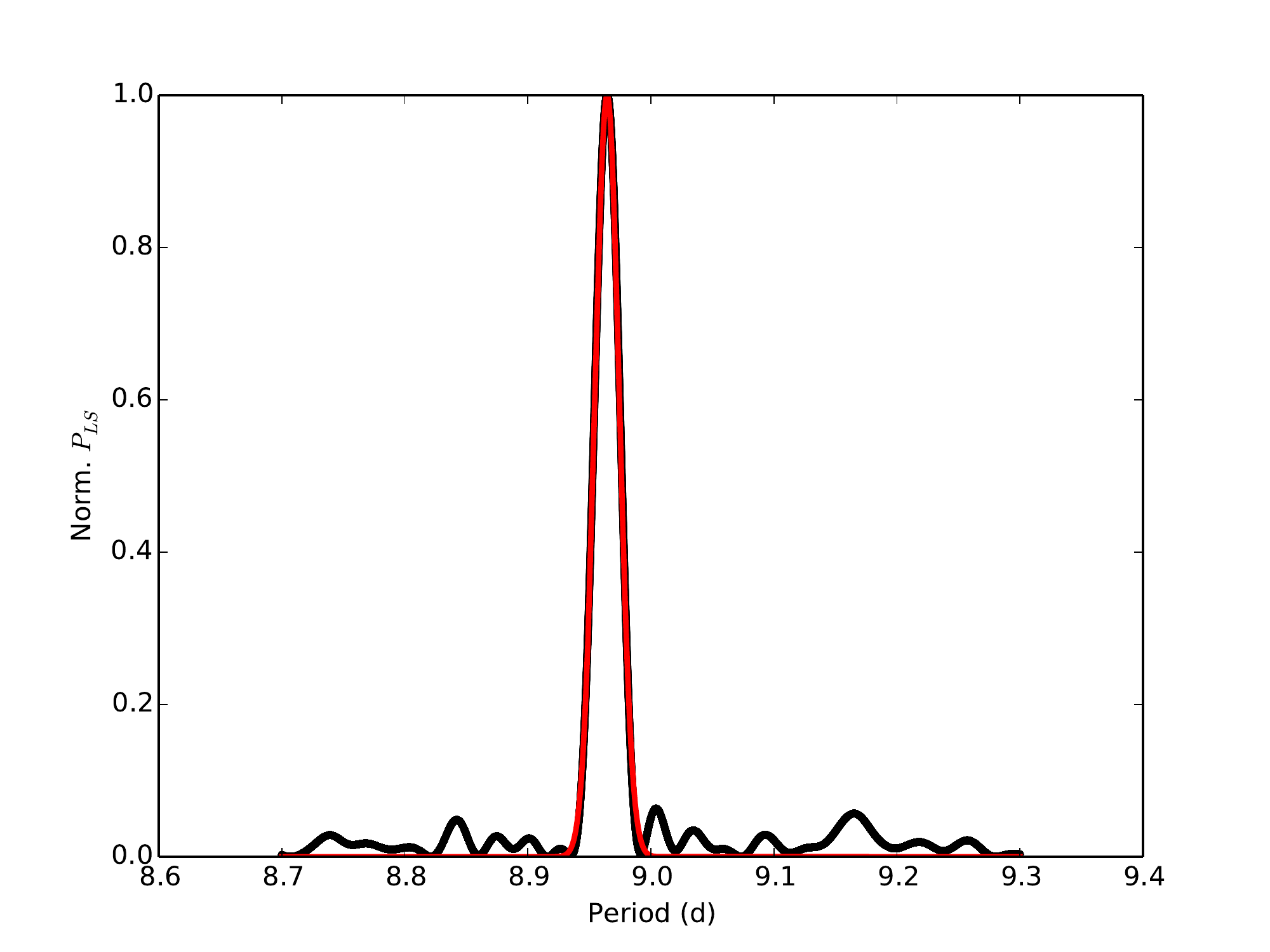}
\caption{The Lomb-Scargle periodogram based on the Swift/BAT 14 -- 100 keV energy band lightcurve. The red line shows the best derived orbital period,  $P=8.964 \pm 0.003$ days.}
\label{fig:LSBAT}
\end{figure}

We have used the Lomb-Scargle \citep{LS_Press} technique to determine the orbital period from the 14 -- 100 keV Swift/BAT light-curve. The orbital period derived from the complete data-set is P$_{orb}$=$8.964\pm0.003$ days.  Figure \ref{fig:LSBAT} shows the Lomb-Scargle power (black line), 
together with a Gaussian fit (in red). 
We searched for variations of the orbital period splitting the data in few equally spaced time intervals. The results are listed in Table \ref{tab:orbPeriod}. The orbital period remains constant within the uncertainties, and no secular evolution can be identified. 
The average period together with the mid-eclipse reference time MJD = 53377.3964 have been used to obtain folded light-curves described below. 
 
\begin{table}
\caption{Orbital periods and uncertainties.}
\centering                          
\begin{tabular}{cc}
\hline                 
\hline                 
\noalign{\smallskip}
Time interval  (MJD) &Period (d)         \\
\noalign{\smallskip}
\hline                 
\noalign{\smallskip}
53370 - 56290 	     &$8.964\pm0.003$ \\
\noalign{\smallskip}
\hline
\noalign{\smallskip}
53370 - 54830	     &$8.967\pm0.007$ \\
54830 - 56290	     &$8.965\pm0.007$ \\
\noalign{\smallskip}
\hline
\noalign{\smallskip}
 53370 - 54344        &$8.97\pm0.01$ \\
 54344 - 55315    	 &$8.96\pm0.01$ \\
 55315 - 56290    	 &$8.97\pm0.01$ \\
\noalign{\smallskip}
\hline              
\end{tabular}
\label{tab:orbPeriod}
\end{table}

\begin{figure}
\centering 
\includegraphics[width=0.5\textwidth]{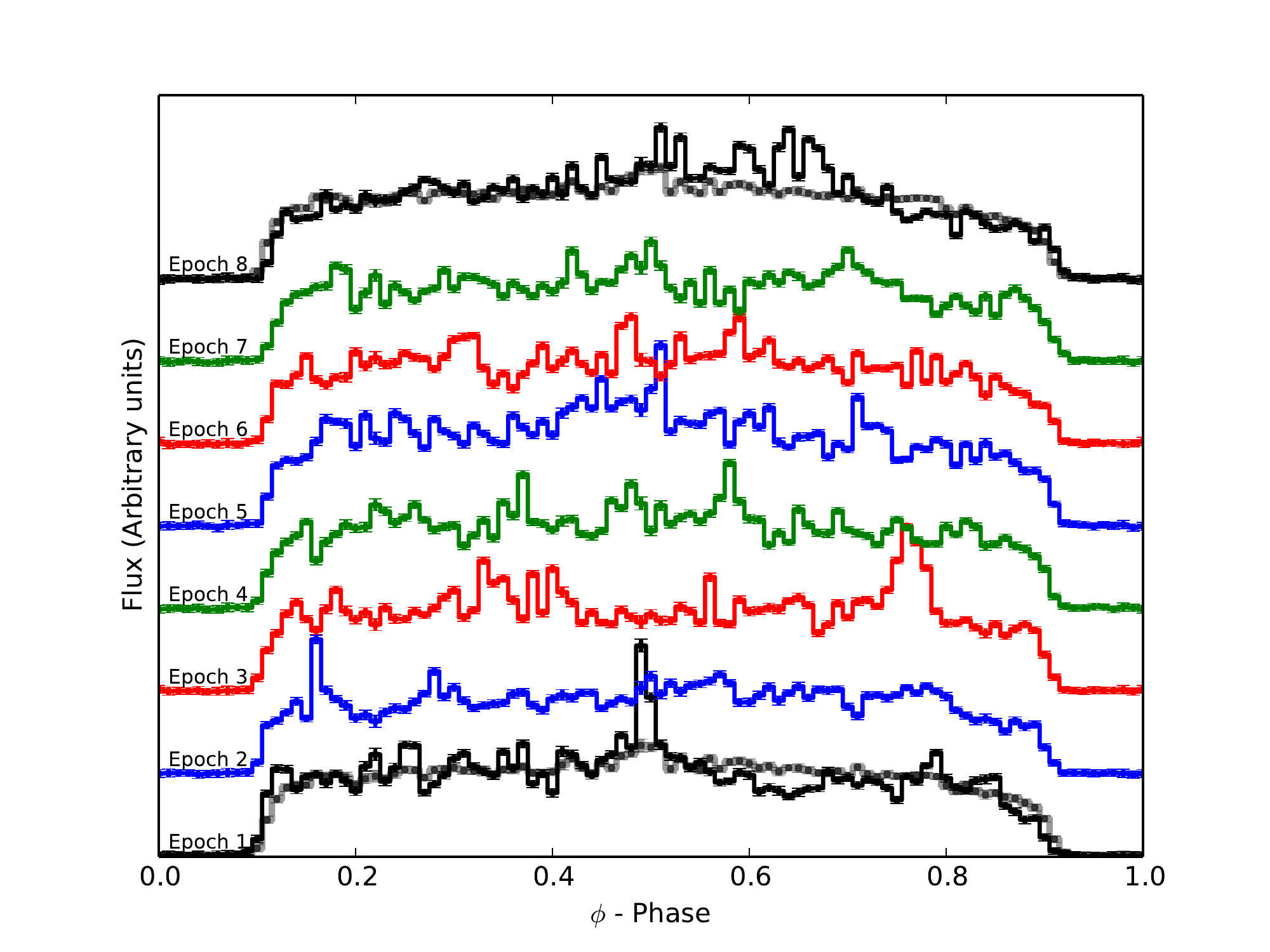}
\caption{ The folded lightcurve of Vela X-1 in the 14 -- 100 keV energy band, split in 8 consecutive datasets
and shifted arbitrarily. The grey curves shown on the top and bottom include all available data for comparison. }
\label{fig:epochs}
\end{figure}

The 14 -- 100 keV lightcurve was also split in 8 successive periods (each spanning about a year) in order to study variability over these periods (Fig. \ref{fig:epochs}). Some major flaring activity can be identified in each of the periods. Vela X-1 is known to be highly variable at hard X-rays and features major flares regularly \citep{Kreykenbohm+08}. These flares (when averaged over a year) are not located at specific orbital phases, indicating that they are stochastic and much shorter than the orbital period. 

\begin{figure}
\centering 
\includegraphics[width=0.5\textwidth]{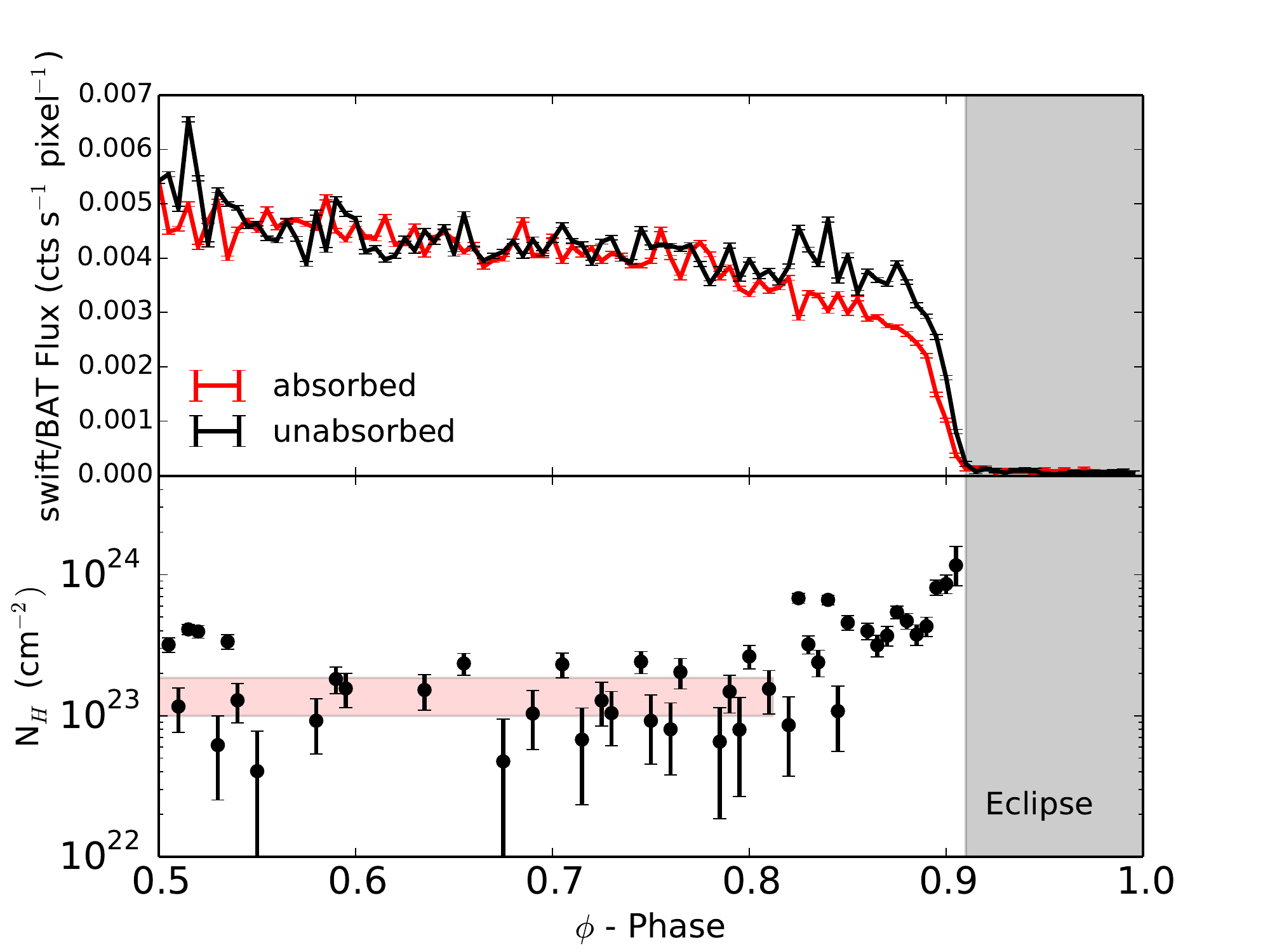}
\caption{{\it Top panel}: The observed folded light-curves preceding (black) and following (red) the eclipse. {\it Bottom panel}: inferred column density. 
The shaded vertical area indicates the eclipse. The light red horizontal area shows the artificial column density variability induced by the source short term intensity variability.}
\label{fig:Assymetry}
\end{figure}

The averaged folded lightcurve shows a clear asymmetry before and after the eclipse (Fig. \ref{fig:Assymetry}). This asymmetry is a likely signature of the accretion wake trailing the neutron star \citep{Blondin90,Blondin91}. Several sgHMXBs \citep[and references therein]{Manousakis+12} show a similar behaviour. 
The average absorbing column density of the accretion wake can be reconstructed from the ratio of the folded lightcurves observed on both sides of the eclipse. In Fig. \ref{fig:Assymetry}, we compare the lightcurve preceding ($\phi=0-0.5$; black line) and following ($\phi=0.5-1$; red line) the eclipse.
For the phases of $\phi\sim 0.5-0.8$ the variations between the two curves are of the order of $(1-2)\times 10^{23}$ cm$^{-2}$ and related to source flares and to inhomogeneities in the stellar wind. A very strong flare, which occurred during the first year of observation, shows up close to phase 0.5. Shortly before ingress (i.e., $\phi\sim 0.8-0.9$) the flux is significantly lower, corresponding to scattering in an additional absorbing column density of the order of $8\times 10^{23}$ cm$^{-2}$. 

The BAT folded lightcurve during eclipse ingress and egress has an exceptional signal to noise. Besides constraining the accretion wake, the lightcurve is particularly sensitive to the velocity of the unperturbed stellar wind close to the surface of the supergiant. An aspect that we can further probe comparing the observations to the results of simulations.

\section{Hydrodynamic Simulations} 
\label{sec:hydro}

The hydrodynamic simulations are largely based on our previous work \citep{Manousakis_offstates} but tuned to the new observational results presented here. Only a brief description of the simulation code is provided here. The Euler equations dictate the motion of the wind, assuming mass, momentum, and energy conservation and account for the Roche potential and the line driven force \citep{Blondin90,Blondin91}.The mesh is fixed but non-uniform, allowing high resolution around the neutron star.

The winds of hot massive stars are characterized observationally by the wind terminal velocity and the mass-loss rate. Wind terminal velocities and mass-loss rates are typically in the range $\upsilon_{\infty}\sim 1000-3000$ km s$^{-1}$ and $\dot{M}_{\rm w}\sim 10^{-(6-7)}$ M$_{\odot}$ yr$^{-1}$, respectively \citep{winds_from_hot_stars}. The wind velocity profile is described by the $\beta$-velocity law $\upsilon=\upsilon_{\infty}(1-R_{*}/r)^{\,\beta}$, where $\beta$ is the velocity gradient \citep{CAKwind}. The values of $\upsilon_{\infty}$ and $\beta$ are used to derive the CAK parameters characterizing the line driven force in our simulations using a 1-D simulation. These CAK parameters are then used in the 2-D code.

The ionization of the wind by the neutron star is characterized by the ionisation parameter $\xi= L_{X}/n r_{ns}^2$,  where $L_{X}$ is the average X-ray luminosity and  $n$ is the number density at the distance $r_{ns}$ from the neutron star \citep{1969ApJ...156..943T}. Close to the neutron star, where $\xi>\xi_{crit}\sim 10^{2.5}$ erg cm sec$^{-1}$, the ions responsible for wind acceleration are highly ionized \citep{Kallman82} and the radiative acceleration cuts-off.  Although, the  effects of the X-ray feedback on the wind acceleration force are complicated due to the large number of ions and line transitions contributing to the opacity \citep{1982ApJ...259..282A,1990ApJ...365..321S} a simplified approach allows the formation of a realistic shock in front of the neutron star \citep{FarnssonFabian1980}. 

We have made use of  VH-1\footnote{http://wonka.physics.ncsu.edu/pub/VH-1/} code, described in depth by \citet{Blondin90,Blondin91}. 
The computational mesh employed in this study consists of 900 radial by 347 angular zones, extending  from 1 to $\sim$ 25 R$_{*}$ and in 
angle from $-\pi$ to $+\pi$. The non-uniform grid is nested around the neutron star, reaching its highest resolution there, about $\sim 10^{9}$ cm . 
A co-rotating reference system located around the center of mass is used. 
The outer boundary condition is characterized as an outflow while an absorbing boundary conditions \citep{1971MNRAS.154..141H,2009ApJ...700...95B}
is used at the stellar photosphere. 
The initial setup (i.e., wind density, velocity, pressure as well as the CAK parameters) is described in \cite{Manousakis_offstates} and
results to $\beta\approx 0.5 - 0.8$ when the 2-D simulation stabilizes.
  
We performed several simulation runs with a wind terminal velocity of $\upsilon_{\infty}\approx 1400,\, 1700$ km s$^{-1}$, $\beta=0.5,\, 0.8$, and binary separations $\alpha=1.76,\, 1.77,\, 1.78\, {\rm R}_{*}$. The mass loss rate of the donor star was always fixed to $\dot{\rm M}\approx 4\times 10^{-6}\, {\rm M}_{\odot}$.  
The time step of the simulations is $\sim 1/10$ sec. The simulations ran for roughly $\ga$ 3 orbital periods. The first $\sim$ 3 days were excluded from the analysis, allowing the wind to reach a steady configuration. Assuming a mass to light conversion factor $\eta=0.1$, the resulting time-averaged mass accretion rate corresponds to an X-ray luminosity in the range $(2-8)\times 10^{36}$ erg/s.

\begin{figure}
\centering 
\includegraphics[width=0.5\textwidth]{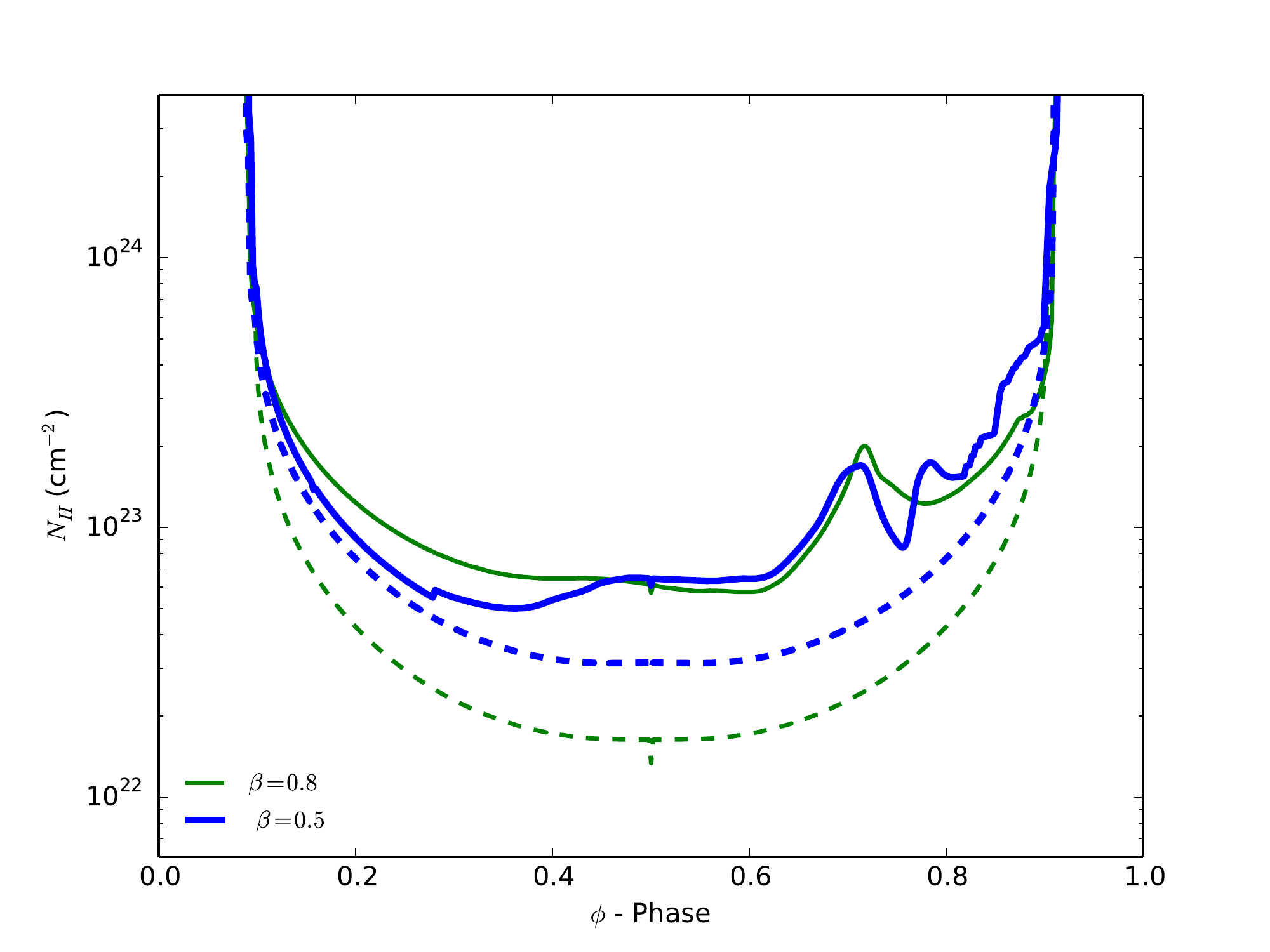}
\caption{The column density as a function of orbital phase derived from a smooth wind (dashed lines) and hydrodynamic simulations  
(solid lines) with a wind terminal velocity, $\upsilon_{\infty}\approx 1400$ km s$^{-1}$ (derived from 1-D simulation) and $\beta=0.5$ (blue) and $\beta=0.8$ (green). }
\label{fig:allNH}
\end{figure}

\begin{figure*}
  \centering
  \subfloat[Matching the observations] {\label{fig:good} 
  \includegraphics[width=0.5\textwidth]{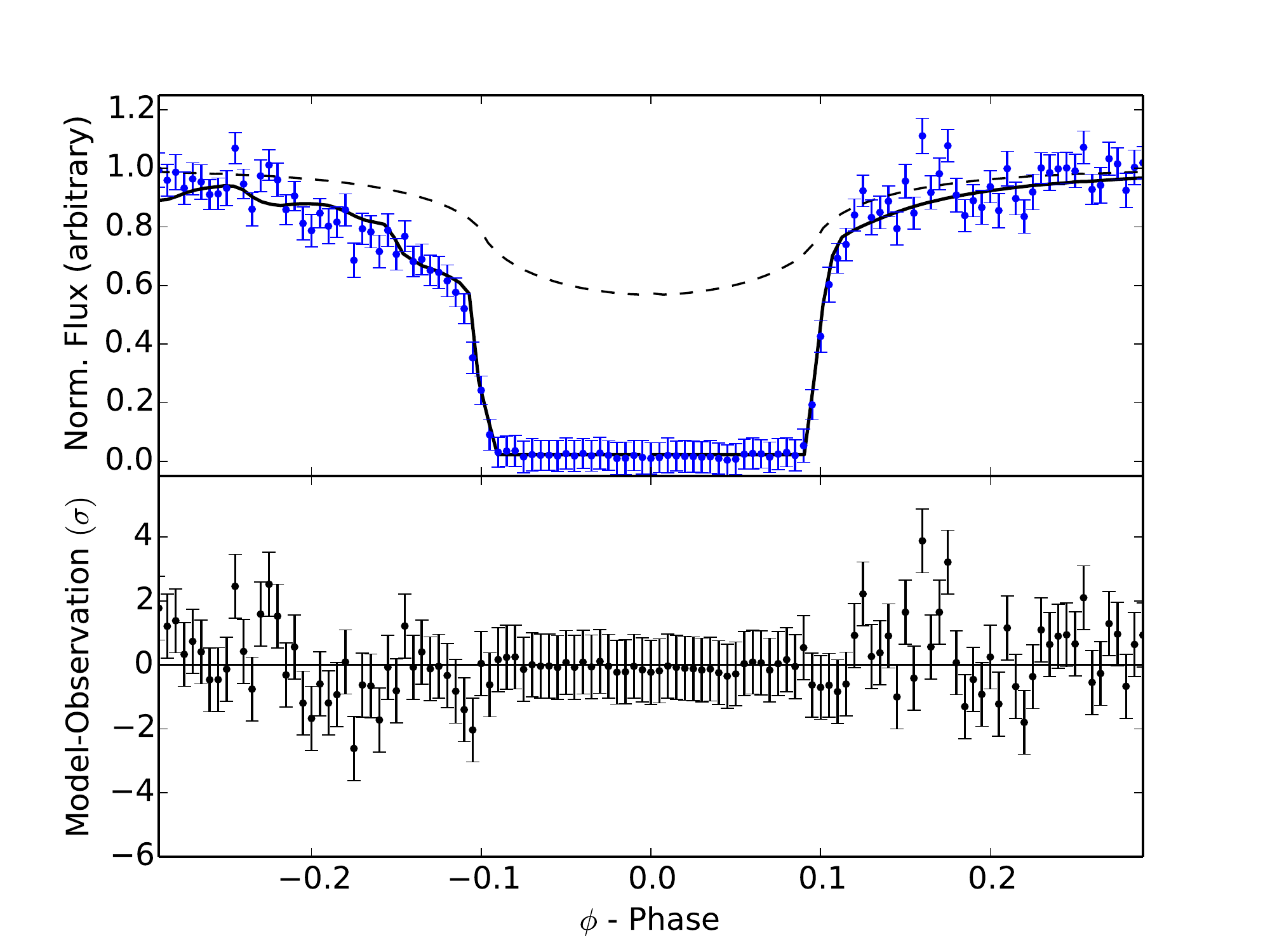}
  }  
  \subfloat[Ruled out by the observations] {\label{fig:bad} 
  \includegraphics[width=0.5\textwidth]{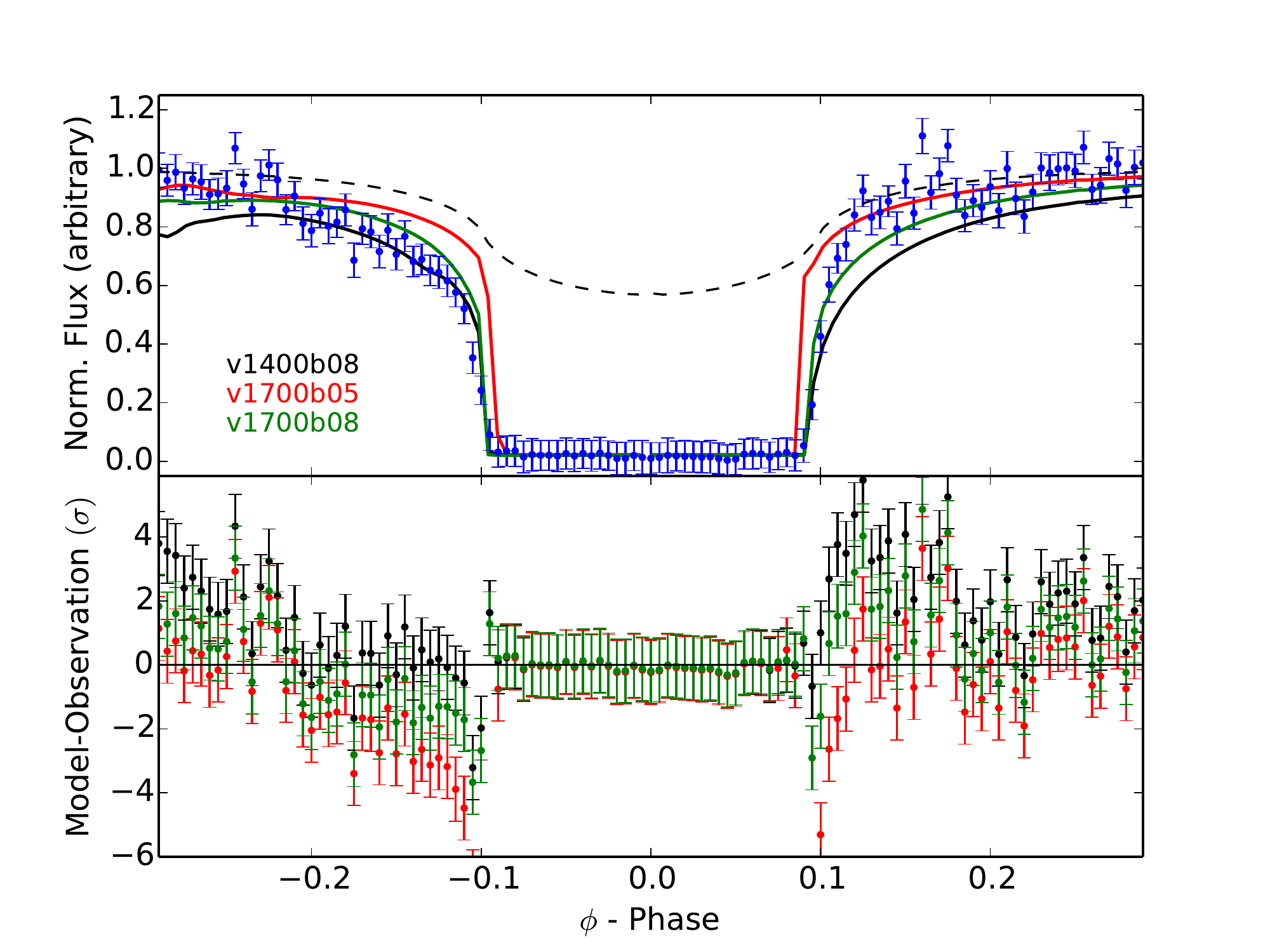}
  }  
\caption{The Swift/BAT 14 -- 100 keV energy range folded light-curves of Vela X-1 (blue points) together with the simulated eclipse profiles (dashed lines) using 
a binary separation $\alpha=1.77$ R$_{*}$. \emph{Left:}  
The light-curves corrected for scattering (solid line) were built assuming wind terminal velocity of $\upsilon_{\infty}=1400$  
and $\beta=0.5$. \emph{Right:} The same as previous for the remaining models, vXXXXbYY, where XXXX is the wind terminal velocity in km s$^{-1}$ and YY is 
the wind $\beta$ parameter. The residuals between the model and observations are shown in the bottom panel. The $\chi^2$ are listed in Table \ref{tab:bestmodel}.}
\label{fig:all_absLC}
\end{figure*}

To compare the observations with the simulations we have built synthetic lightcurves derived for each simulation run. The neutron star is emitting as a point source and part of these X-ray photons  are scattered in the stellar wind forming a diffuse source. The global intensity and geometry of the scattered emission was calculated from the illumination and Thomson optical depth of each simulation cell. This is a reasonable assumption at hard X-rays. We assumed that the accretion wake extends vertically in the range $|z|$=$0.5\, R_{*}$ and that above these limits a smooth stellar wind applies. The very small flux (with a relative emissivity of $10^{-4}$) observed during eclipse originates from an extended region of the wind and was accounted for. These synthetic lightcurves are shown as dashed lines in Fig. \ref{fig:all_absLC}. 

We also calculated the column density ${\rm N}_{\rm H}$ between the observer and the neutron star along the orbit cell by cell, excluding the highly ionised part of the wind, and corrected the synthetic lightcurve for the effect of scattering. These lightcurves are displayed as continuous lines in Fig. \ref{fig:all_absLC} together with the Swift/BAT data (blue points). The correspondence between the simulations and the data, measured with a reduced $\chi^{2}$, are listed in Table \ref{tab:bestmodel}. An excellent match is obtained for a steep ($\beta=0.5$) velocity law and a terminal velocity of 1400 km/s. These values were the inputs to the 1-D simulation used to derive the CAK parameters and correspond to an unperturbed wind velocity of $\approx 1400$ km/s at the orbital radius and of 850 km/s at a distance of 0.2 R$_*$ of the stellar surface in the 2-D simulations. Binary separation significantly different from $\alpha=1.77\, {\rm R}_{*}$ induces large variations at eclipse ingress and egress that do not match the observed data ($\chi^{2}\ga 200$ for 119 d.o.f.). 

The time-averaged absorbing column density derived from the simulations are shown in  Fig. \ref{fig:allNH} for $\beta=0.8$ (green) and $\beta=0.5$ (blue) together with the predictions from an unperturbed smooth stellar wind (dashed lines). Significant ${\rm N}_{\rm H}$ variability is expected from orbit to orbit.  
 The average Swift/BAT spectrum for that phase range is not sensitive to low column densities so could not be used to further constrain the model parameters.


\begin{table}
\caption{Simulation parameters (wind terminal velocity, beta parameter and binary separation) and agreement ($\chi^{2}$) with the observed lightcurve.}
\centering                          
\begin{tabular}{lccc}
\hline                 
\hline                 
\noalign{\smallskip}
Model & $\upsilon_{\infty}$ (km s$^{-1}$) & $\beta$ & $\chi^{2}$/d.o.f         \\

\noalign{\smallskip}
\hline                
\noalign{\smallskip}
$\alpha$=1.76 R$_{*}$ & & &  \\ 
\hline
\noalign{\smallskip}
\verb=v1700b08= &1700& 0.8 & 407 / 119\\
\noalign{\smallskip}
\verb=v1400b08= &1400& 0.8 & 382 / 119\\
\noalign{\smallskip}
\noalign{\smallskip}
\verb=v1700b05= &1700& 0.5 & 392 / 119\\
\noalign{\smallskip}
\verb=v1400b05= &1400& 0.5 & 212 / 119\\
\noalign{\smallskip}
\hline                
\noalign{\smallskip}
$\alpha$=1.77 R$_{*}$ & & &  \\ 
\noalign{\smallskip}
\hline                 
\noalign{\smallskip}
\verb=v1700b08= &1700& 0.8 & 297 / 119\\
\noalign{\smallskip}
\verb=v1400b08= &1400& 0.8 & 276 / 119\\
\noalign{\smallskip}
\noalign{\smallskip}
\verb=v1700b05= &1700& 0.5 & 295 / 119\\
\noalign{\smallskip}
\verb=v1400b05= &1400& 0.5 & 128 / 119\\
\noalign{\smallskip}
\hline
\noalign{\smallskip}
$\alpha$=1.78 R$_{*}$ &  & &  \\ 
\hline
\noalign{\smallskip}                 
\noalign{\smallskip}
\verb=v1700b08= &1700& 0.8 & 452 / 119\\
\noalign{\smallskip}
\verb=v1400b08= &1400& 0.8 & 405 / 119\\
\noalign{\smallskip}
\noalign{\smallskip}
\verb=v1700b05= &1700& 0.5 & 399 / 119\\
\noalign{\smallskip}
\verb=v1400b05= &1400& 0.5 & 253 / 119\\
\noalign{\smallskip}

\hline              
\end{tabular}
\label{tab:bestmodel}
\end{table}

\section{Discussion}   
\label{sec:disc}

Spectroscopic observations of massive stars allow to constrain the mass loss rates and the terminal velocities of stellar winds \citep[and references therein]{Puls+08Review}. The P-Cygni line profiles used for these estimates are formed far from the massive star at a few tens of stellar radii, where the $\beta$ parameter has a small effect. Orininally the models by \citet{CAKwind} resulted in a steep $\beta=0.5$ law, while later on a shallower $\beta=0.8$ was introduced to match the observations \citep{1986A&A...164...86P,1986ApJ...311..701F,Puls+08Review}.  \cite{1993MNRAS.265..601S} discussed the difficulties of low velocity gradients and the need for gradients varying with radius. In an attempt to study the stellar winds from Wolf-Rayet stars, \citet{1994A&A...289..505S} employed a modified $\beta$ law mimicking a steep velocity field gradient ($\beta=0.5$) close to the stellar surface and a smoother ($\beta=1.0$) velocity gradient further away. Such a low $\beta$ parameter close to the stellar surface was also found studying the soft X-ray absorbing column density along the orbit in 4U 1700-37 with {\it EXOSAT} \citep{1989ApJ...343..409H} but these measurements were much more sensitive to ionisation than ours, based on hard X-ray data. 

The line driven instability  predicts large density and velocity discontinuities which are not modelled in our simulations. According to the latest models \citep{2013MNRAS.428.1837S}, this occurs at distances larger than one stellar radius from the stellar surface, i.e. well outside of the orbit in the case of Vela X-1, and should not play an important role in driving the X-ray variability. Gravitation and the X-ray ionisation from the neutron star, that are taken into account in our simulations, remain the main drivers of the hydrodynamics. Photoionisation influences the wind within a few tens of the orbital radius \citep[see also][]{2012ApJ...757..162K} but leaves the wind located at $\pm 90^\circ$ from the neutron star direction, responsible for the inress/egress scattering profile, largely unaffected.

The hard X-ray scattering profile measured at eclipse ingress and egress is sensitive to the density of the stellar wind at a distance of a few tenths of stellar radius from the stellar surface. At eclipse ingress it is in addition sensitive to the column density of the accretion wake formed close to the orbital radius. The density of the wind depends directly on the $\beta$ parameter. A larger velocity gradient will generate a longer eclipse egress. Note that the wind velocity at 1.2 R$_*$ is about twice larger with $\beta=0.5$ than with $\beta=0.8$ and has therefore a strong impact on the hard X-ray flux profile. The column density of the accretion wake also depends on the $\beta$ parameter, a slower wind at the orbital radius will be more strongly deflected by the neutron star and form a stronger wake, i.e. a longer and more structured eclipse ingress. 

\begin{figure}
  \centering
  \includegraphics[width=0.5\textwidth]{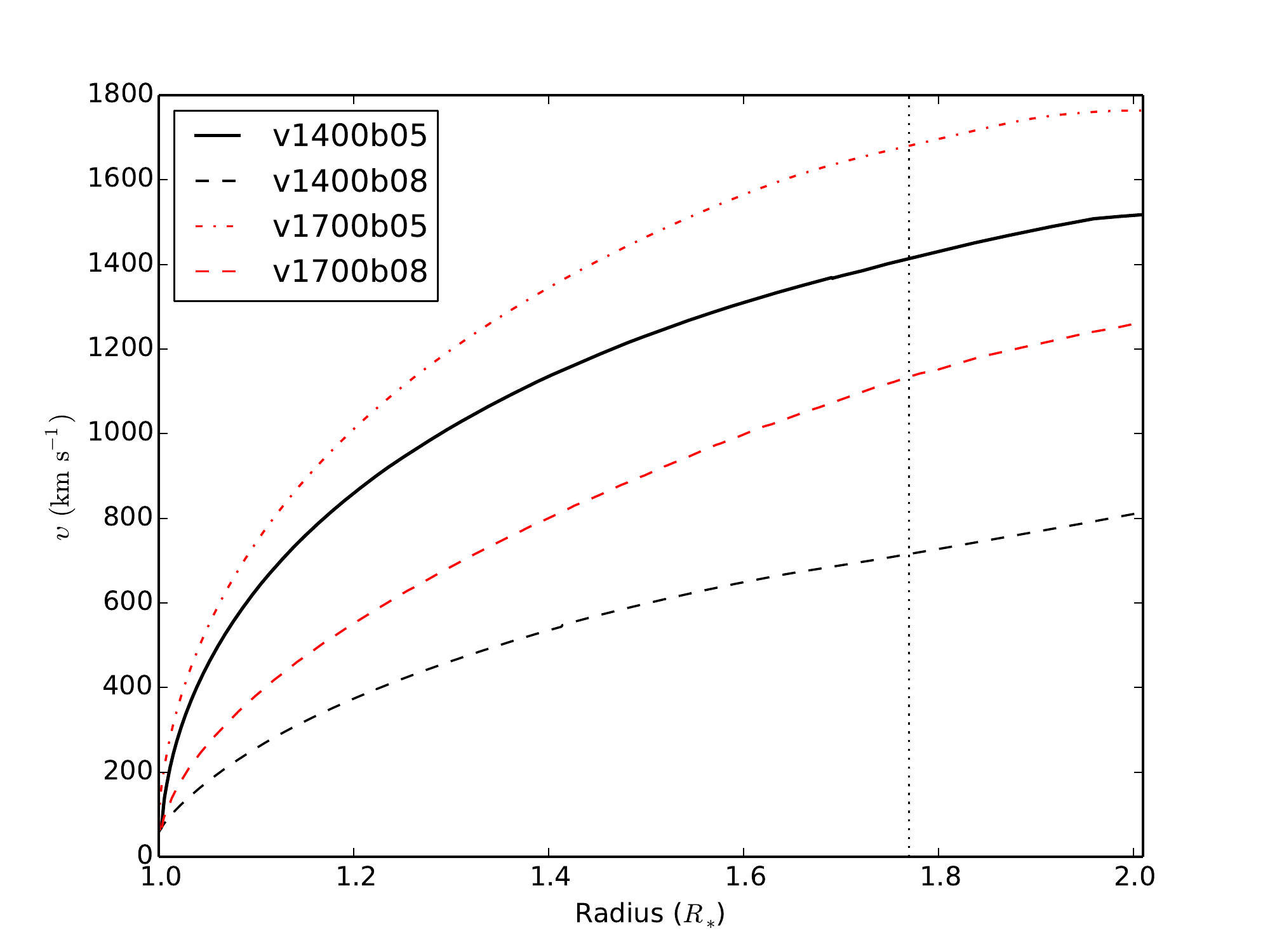}
\caption{Wind velocity field close to the surface (R$_*=1$) of the supergiant. The curve labelled as \texttt{v1400b05} provides an excellent agreement with the observations. The other velocity fields do not provide results consistent with the observations. The vertical dotted line indicates the orbital radius.}
\label{fig:vel}
\end{figure}

Our simulation provides an excellent match for both the egress and the ingress hard X-ray lightcurves with a single set of parameters ($\upsilon_{\infty}$=1400 km/s and $\beta=0.5$, from the 1D simulation), which is remarkable. Fig. \ref{fig:vel} shows the best fit local velocity derived from the 2-D simulations in the unperturbed wind (at 90$^\circ$ from the direction to the neutron star). This 2-D velocity profile can be represented with the usual $\beta$ law with the parameters $\beta=0.55$
and $\upsilon_{\infty}$=1700 km/s, the latter is in agreement with the measurements from \cite{1980ApJ...238..969D}.
The simulated folded lightcurve is very sensitive to the velocity profile and constrain the velocity within about $\pm 100$ km/s at 1.2 R$_*$. 

It is important to note that the X-ray data are not sensitive to the velocity of the wind beyond the orbital radius, so a larger observed $\upsilon_{\infty}$ can be accommodated if the $\beta$ parameter increases towards larger distances. 
In the past the effects of the accretion wake have been studied in detail by \cite{Blondin90} but not related to the variability at egress. More recently
\cite{2012MNRAS.421.2820O} used a highly structured clumpy stellar wind to explain soft X-ray absorption along the orbit, resulting in strong variability and a symmetry around the eclipse, contrasting with the data and the likely absence of clumps very close to the stellar surface.

\section{Conclusion}   
\label{sec:conclude}
 
We have analysed almost a decade of $Swift$/BAT data of Vela X-1. Although variable, its hard X-ray emission, folded along the orbit, features a persistent asymmetric eclipse ingress and egress, measured with high signal-to-noise. This asymmetry originates from the scattering of hard X-rays around the neutron star at small and larger scales, including in the accretion wake. The X-ray variability is related to in-homogeneities in the stellar wind largely created by the gravity and ionisation of the neutron star itself \citep{Manousakis_offstates}.  

The simulation matches the observed data for a narrow set of parameters implying that the wind velocity close to the stellar surface is twice larger than usually assumed. The stellar wind in Vela X-1 shows velocity field similar to these observed in some Wolf-Rayet stars \citep{1994A&A...289..505S}.

The hydrodynamic simulations of the wind of Vela X-1 are still incomplete, however they provide already an excellent agreement with the observations for the smooth properties (this work) and the variability \citep{Manousakis_offstates}. High-mass X-ray binaries provide informations on the initial acceleration of the stellar wind, which are unique and required to test models accurately.

\begin{acknowledgements}
This work was partially supported by the Polish NCN grants 2012/04/M/ST9/00780 and 2013/08/A/ST9/00795. 
\end{acknowledgements}

\bibliographystyle{aa} 
\bibliography{references} 

\end{document}